\documentclass[aps,prb,twocolumn,groupedaddress,showpacs,superscriptaddress,amssymb,amsmath]{revtex4-1}
\usepackage[font=footnotesize,labelfont=bf,   justification=raggedright, format=plain]{subcaption} 
\usepackage{mathtools}
\usepackage{amsmath}
\usepackage{graphicx}
\usepackage{color}
\usepackage[font=footnotesize, labelfont=bf,   justification=raggedright,  format=plain]{caption} 
\usepackage{tabularx}
\usepackage{comment}
\newcommand{\be}{\begin{equation}}
\newcommand{\ee}{\end{equation}}
\newcommand{\bea}{\begin{eqnarray}}
\newcommand{\eea}{\end{eqnarray}}
\usepackage{dcolumn}
\usepackage{hyperref}
\usepackage{bm}
\usepackage{epstopdf}
\usepackage{amsfonts}

\begin{document}

\title{Phononic heat transport in molecular junctions: quantum effects and vibrational mismatch}

\author{Roya Moghaddasi Fereidani}
\affiliation{Chemical Physics Theory Group, Department of Chemistry, and
Centre for Quantum Information and Quantum Control,
University of Toronto, 80 Saint George St., Toronto, Ontario, Canada M5S 3H6}
\author{Dvira Segal}
\affiliation{Chemical Physics Theory Group, Department of Chemistry, and
Centre for Quantum Information and Quantum Control,
University of Toronto, 80 Saint George St., Toronto, Ontario, Canada M5S 3H6}

\date{\today}

\begin{abstract}
Problems of heat transport are ubiquitous to various technologies such as power generation,
cooling, electronics, and thermoelectrics. 
In this paper we advocate for the application of the quantum self-consistent reservoir method, 
which is based on the generalized quantum Langevin equation,
to study phononic thermal conduction in molecular junctions. 
The method emulates phonon-phonon scattering processes while taking into account quantum 
effects and far-from-equilibrium (large temperature difference) conditions. 
We test the applicability of the method by
simulating the thermal conductance of molecular junctions with
one-dimensional molecules sandwiched between solid surfaces.
Our results satisfy the expected behavior of the thermal conductance 
in anharmonic chains as a function of length, phonon scattering rate and temperature, 
thus validating the computational scheme. 
Moreover, we examine the effects of vibrational mismatch between the solids' phonon spectra on 
the heat transfer characteristics in molecular junctions. 
Here, we reveal the dual role of vibrational anharmonicity: 
It raises the resistance of the junction due to multiple scattering processes, 
yet it promotes energy transport across a vibrational mismatch by enabling phonon 
recombination and decay processes.
\end{abstract}

\maketitle


\section{Introduction}
\label{sec-intro}

With the emergence of single-molecule electronics \cite{MolElRev}, 
the related problem of thermal transport through molecular junctions (MJs) 
has received considerable theoretical and experimental attention 
\cite{GalperinRev,DubiRev,HanggiRev,SegalRev,ReddyRev}.  
In metals, heat is mainly carried by electrons, whereas in semiconductors and electrical insulators 
it is primarily transmitted by lattice vibrations, termed phonons \cite{Ashcroft}.  
Down to the nanoscale, thermal management in low dimensional systems is a topic 
of a significant technological importance
\cite{CahillRev1,CahillRev2,Gang}.
In molecular junctions made of poorly electron-conducting components
such as saturated hydrocarbons, excited phonon modes in the contacts 
(solids) exchange energy with molecular vibrations;
thermal energy is then carried away 
through different atomic motions in the molecule, e.g. stretching modes 
\cite{Segalmanexp,Dlott,Gotsmann,RubtsovRev,RubtsovAcc}. 

Undoubtedly, quantum effects should be taken into account when studying vibrational 
heat transport through molecules since the frequencies of molecular vibrations
oftentimes lie above the thermal energy \cite{LeitnerRev1,LeitnerRev2,Straub,DharRev,LuRev,BijayRev}. 
Anharmonic interactions 
significantly impact the thermal conductivity of bulk materials and nanostructures at room temperatures.
Most notably, anharmonic interactions lead to inelastic phonon-phonon scattering processes, thermalization, and
ultimately the development of the Fourier's law of thermal conductivity in macroscopic objects.  
Likewise, in small molecules anharmonic interactions 
are critical to enforce intra- and inter-molecular vibrational redistribution (IVR) processes, essential 
in reaction dynamics \cite{LeitnerRev1,Straub,RubtsovRev}, as well as in
nonlinear and nonreciprocal effects  \cite{SegalRev,dlottU1,dlottU2,HanggiRev}.

Performing quantum simulations of phononic heat transport in molecules---while taking into 
account anharmonic interactions---is a formidable task and a 
topic of a significant research work 
\cite{DharRev,LuRev,BijayRev,SegalRev,LeitnerRev1,LeitnerRev2,ReddyRev}. 
It is feasible to employ anharmonic force fields when using classical molecular dynamics simulations \cite{Lepri,Segal08,Yun10}, 
but such calculations are limited to the high temperature regime when quantum effects are not influential.
From the other end, full quantum calculations 
can be performed based on the Landauer formalism, once assuming
a complete harmonic model \cite{Kirczenow,Segal03,DharRev,Pauli,Gemma1,Cuevas17,Cuevas}.
Among the quantum-based methods for molecular thermal conduction, which take into account anharmonic interactions,
we recall the nonequilibrium Green's function technique \cite{MingoRev,BijayRev,Thoss10,Nitzan07}
as well as kinetic approaches  \cite{LeitnerRev1,Straub,Segal06,Juzar12};
both methods are limited to systems with few quantum states and 
to models with a perturbative parameter (weak system-bath coupling or small anharmonicity). 
Other techniques are based e.g. on the Born-Oppenheimer principle \cite{Wu11},
mixed quantum classical propositions \cite{WangSC,joe}, and the renormalization of normal modes \cite{Junjie1,Junjie2}.
Numerically exact simulations in the language of wavefunctions \cite{ThossCPL} or 
path integral representations 
\cite{Saito13, SaitoNJP,Segal13,Nazim}
are limited to minimal models such as the nonequilibrium spin-boson model.

\begin{figure}
\includegraphics[scale=0.45]{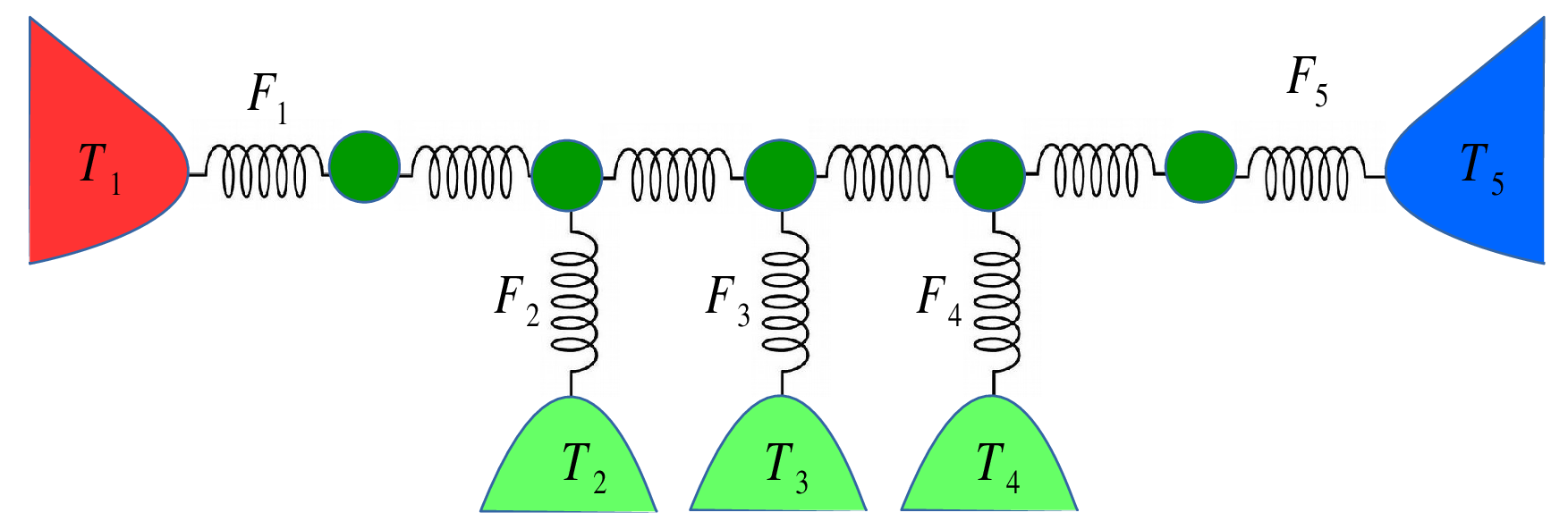} 
\caption{Scheme of a chain with $N=5$ particles (filled circles);
springs represent harmonic bonds.
The temperatures $T_1$ and $T_5$ define the boundary condition, while
the inner baths' temperatures $T_{2, 3, 4}$ are determined such that the 
individual heat currents towards these baths, $F_{2, 3, 4}=0$, vanish.
The net heat current flowing across the junction is given by $F_1=-F_5$.}
\label{Fig1}
\end{figure}

The central objective in this paper is to test and analyze a practical atomistic tool for studying
phononic heat transfer including quantum mechanical effects
in anharmonic MJs.
Our method of choice is the quantum self-consistent reservoir (QSCR) approach,
which can smoothly interpolate between the ballistic and diffusive transport regimes
\cite{Roy06, Roy08,DharRev,Malay11,nanoSCR}.
In this technique,
phonon-phonon scattering processes, the outcome of the anharmonicity of the inter-atomic potential, 
are introduced phenomenologically-operationally. 
Specifically, particles (atoms or coarse grained entities) are attached to 
inner, self-consistent (SC) heat reservoirs that serve to mimic the role of inter-particle anharmonic interactions. 
Fig. \ref{Fig1} displays the QSCR model for a linear chain bridging
two solids that are maintained at different temperatures.
Each internal site is coupled to an independent heat bath, and the temperatures of the inner baths
are determined by enforcing the condition that, in
steady-state, on average there is no net heat flow between interior heat baths and the junction.
This condition ensures energy conservation in the physical system---without conserving the 
phonon number--- thus emulating phonon-phonon scatterings  in the molecule.

The classical version of the QSCR method was proposed in Refs. \cite{SCR70,SCR75}. 
More recently, the model was solved exactly in the thermodynamic limit in Refs. \cite{Bonetto04,Pereira06},  
manifesting the development of local equilibrium and the onset of the Fourier's law of
heat conduction.  
The quantum version of this model was studied by Visscher and Rich \cite{SCR75S}, who 
analyzed the case of weak system-bath coupling. 
The quantum model was later reconsidered by Dhar and Roy \cite{Roy06,Roy08} beyond the weak coupling limit. 
More recently, Bandyopadhyay and Segal \cite{Malay11} applied an iterative numerical procedure and used the 
QSCR model beyond the linear-response approximation 
so as to examine the phenomenon of thermal rectification in short MJs. 
Massive, atomistic QSCR simulations of heat transport in graphene nanoribbons 
were performed in Ref.\cite{nanoSCR}.

While the literature over the classical and quantum self-consistent reservoir method
is quite extensive, the method was not assessed
for finite-size, physical MJs (unlike the thermodynamic limit of infinitely long chains,
which is interesting for fundamental questions \cite{Bonetto04}).
In particular, short junctions may experience large temperature drops, 
thus the (linear response) thermodynamic limit
does not correctly represent their behavior.
With the goal to advocate for the application of the QSCR method to study thermal transport 
in low dimensional nanostructures 
such as single-molecule junctions, self-assembled monolayers and amorphous polymers,
we hereby carefully examine the performance of the method in relatively short molecules of 5-20 units.

Alkane chains with 5-12 carbon atoms are poor conductors of charge carriers
thus one can safely assume that their thermal conductance is determined
by vibrational heat transfer. Using a scanning thermal microscopy, 
the phononic thermal conductance of self-assembled monolayers of
alkanes was recently measured,
demonstrating a non-monotonic behavior with length \cite{Gotsmann}. 
Other experiments further found that the thermal conductance of 
alkane chains was enhanced when the molecules were strongly bonded (chemisorbed) to the metal surface, compared to
the case of weaker (physisorbed) interactions to the surface \cite{Cahill1}. 

A recent experiment particularly explored the involved roles of quantum effects and anharmonicity 
in the thermal conduction of short molecules \cite{Majumdar15}:
Self-assembled monolayers of alkanes with about 10 methylene units were
sandwiched between metal leads with distinct phonon spectra---characterized 
by different Debye frequencies. Considering different solids, it was found 
that the thermal conductance decreased as the Debye-frequency mismatch between the two solids
increased. Nonetheless, quite surprising, classical molecular dynamic simulations displayed 
the {\it opposite trend}. 
To explain this discrepancy, it was pointed out in Ref. \cite{Majumdar15} 
that in classical simulations high frequency modes were activated, 
with thermal population exceeding quantum statistical calculations.
In turn, these active high frequency modes 
enhance phonon-phonon scatterings, promoting heat transport over a vibrational mismatch.

Intrigued by issues raised in the phonon-mismatch experiment 
and the unsettling status of simulations in this area, 
in this paper we employ the QSCR method,  
and study the following question: 
How do quantum effects, anharmonicity, vibrational mismatch and (possibly) large temperature differences
play together to determine the thermal conductance of MJs?
The latter point, having the system operating far from equilibrium, poses a significant challenge yet
technological opportunities for small devices \cite{HanggiRev}.

We focus on one-dimensional (1D) chains and employ the QSCR method. 
When the temperature difference is small, the computational
task is simple: It involves the solution of an algebraic equation. 
Beyond linear response, we solve numerically a set of coupled nonlinear equations by following the 
procedure described in Ref. \cite{Malay11}. 
The model and the QSCR method are discussed in Sec. \ref{sec-model}.

The objectives of this work are twofold: 
(i) To demonstrate that the QSCR method can capture the role of genuine anharmonicity 
in quantum thermal conductance at the nanoscale. 
In Sec. \ref{sec-res1} we therefore study the thermal conductance of MJs
with increasing molecular length, temperature and phonon scattering rate.
(ii) To explore the effect of mismatched phonon spectra at the contacts 
on the thermal conductance of molecular junctions.
In Sec. \ref{sec-res2} we examine the roles of anharmonicity, quantumness 
and phonon mismatch in thermal transport. Our results are discussed in light of the alkane
chain experiments of Ref. \cite{Majumdar15},
but the QSCR method becomes particularly compelling when inelastic, thermalization effects
are paramount, as we discuss and
conclude in Sec. \ref{sec-summ}. 

\section{The self-consistent reservoir model}
\label{sec-model}

We focus on linear chains representing oligomers such as alkane  \cite{Rubtsovalkane}
and polyethylene glycol \cite{RubtsovPEG} molecules 
 and calculate the heat current flowing 
through the junction using the QSCR method. 
The technique can be derived from the generalized quantum Langevin equation by
calculating steady-state properties using the Green's function formalism \cite{DharRev,nanoSCR}. 
To study heat transport far from equilibrium, we employ the numerical method described in Ref. \cite{Malay11}. 
Nevertheless, here we generalize this tool and treat `structured' reservoirs with phonon spectra that
are characterized by a central Debye frequency,
unlike the structureless baths used in Ref. \cite{Malay11}.  
For completeness, next we review the working equations of the QSCR method.

In our setup, a one-dimensional chain with $N$ particles bridges two thermal reservoirs, see Fig. \ref{Fig1}.
The first and last particles are attached to hot and cold baths, with temperatures $T_1$ and $T_N$, respectively.
We refer to these baths, which dictate the boundary condition, as the `physical' reservoirs.
The inner $N-2$ beads are each connected to an independent,
inner heat bath that introduces scattering effects without energy dissipation. 
In what follows, we refer to these inner baths as `self-consistent' (SC) reservoirs. 

The main appeal of the QSCR method is that
we capture the anharmonic part of the inter-atomic
potential using harmonic degrees of freedom.
The entire system, chain (denoted by $S$), left ($L$) and right ($R$) reservoirs, 
as well as the inner ($I$) baths are described by harmonic potentials, 
\bea
{\cal H}={\cal H}_S+{\cal H}_L+{\cal H}_{LS}+{\cal H}_R
+{\cal H}_{RS}+\sum_{I}{\cal H}_{I}+\sum_{I}{\cal H}_{IS},
\nonumber\\
\label{eq:fullH}
\eea
with
\bea
{\cal H}_S&=&
\frac{1}{2} P_{S}^{T} M_{S}^{-1} P_{S} + \frac{1}{2} X_{S}^{T} K_{S} X_{S},\,\,
\nonumber\\
{\cal H}_B&=&\frac{1}{2} P_{B}^{T} M_{B}^{-1} P_{B} + \frac{1}{2} X_{B}^{T} K_{B}X_{B},
\nonumber\\
{\cal H}_{BS}&=& X_{S}^{T} K_{SB} X_{B}.
\label{eq:Hs}
\eea
The index $B= L, R, I$ denotes the different reservoirs, 
$M_S$ and $M_B$ are real diagonal matrices with the masses of the particles 
in the chain and in the baths, respectively. 
The real symmetric matrices $K_S$ and $K_B$ are the 
force-constant matrices of the chain and reservoirs, respectively.
The force-constant coefficients between the chain and the baths are included in $K_{SB}$. 
The column vectors $X_S$ and $X_B$ are Heisenberg operators 
representing the particles' displacements about equilibrium positions with 
corresponding momenta operators $P_S$ and $P_B$. 
The operators satisfy the commutation 
relations $[X_j,P_{j'}]=i\hbar\delta_{j,j'}$ and $[X_j,X_{j'}]=[P_j,P_{j'}]=0$. 
The Heisenberg equations of motion for the chain (in a matrix notation) is
\bea
&&M_S\ddot{X}_S=
\nonumber\\
&&-K_SX_S-K_{SL}X_L-K_{SR}X_R-\sum_{I}K_{SI}X_I.
\label{eq:EOMsystem}
\eea
The displacements of the baths satisfy
\bea
M_B\ddot{X}_B=-K_BX_B-K_{SB}X_S, \qquad B=L, R, I.
\label{eq:EOMbaths}
\eea
Specifically, assuming a one-dimensional chain with nearest-neighbor couplings,
the system Hamiltonian is
\bea
{\cal H}_S=\sum_{s=1}^N\frac{1}{2}m_s\dot{x}_s^2 + 
\sum_{s=1}^{N-1}\frac{1}{2}m_s\omega_0^2(x_s-x_{s+1})^2,
\label{eq:H1Dchain}
\eea
with $m_s\omega_0^2$  as the force-constant between particles. Here $x_s$ is the displacement from equilibrium
of the $s$ particle.

Since the baths are harmonic and their coupling to the system is bilinear,
the dynamics can be exactly described
by the generalized Langevin equation \cite{Roy06,TuckermanB}. 
Briefly, we solve the Heisenberg equations of motion of the reservoirs,
Eq. (\ref{eq:EOMbaths}),
and substitute the solution into Eq. (\ref{eq:EOMsystem}). This procedure yields
\bea
m_s\ddot{x}_s(t)&=&-m_s\omega_0^2\,[2x_s(t)-x_{s-1}(t)-x_{s+1}(t)]
\nonumber\\
&-&\int_{-\infty}^{t}\,d\tau\,\dot{x}_s(\tau)\,\Gamma_s(t-\tau)+\eta_s(t),\,\,\,\,\, s=1, 2, ..., N,
\nonumber\\
\label{eq:langevin}
\eea
which is known as the generalized quantum Langevin equation. 
The integral in this equation is responsible for dissipating energy 
from the system into the environment.
The last term is the fluctuating force of the bath acting on the system.
Note that in our model, every particle is coupled to a heat bath. Therefore,
the index $s$ that labels the friction kernel and the fluctuating force 
uniquely identifies the corresponding (physical or SC) bath.

We write down the dynamical friction kernel in terms of a memory function as
$\Gamma_s(t-\tau)=2\gamma_s\,f(t-\tau)$.
For structureless-memoryless baths, $\Gamma_s(t-\tau)=2\gamma_s\,\delta(t-\tau)$.
We use a Debye model to analyze the role of vibrational mismatch on heat transfer:
The baths are characterized by the frequency $\omega_d$, 
and in frequency domain it is given by,
\bea
\Gamma_s(\omega)=\gamma_s\, e^{-\lvert \omega \rvert/\omega_{d,s}}.
\label{eq:friction}
\eea
Below we distinguish between `structureless baths', with $\omega_d\to \infty$ and `Debye baths',
with  $\omega_d\sim \omega_0$. 
In the Langevin equation description, $\gamma$ is the friction coefficient. Similarly, here
 this coefficient quantifies energy exchange with the baths, but it takes different roles for the physical and internal reservoirs:
The coefficients $\gamma_{L,R}$ represent the bond energy of the molecule to the physical solids.
In contrast,  $\gamma_{i}$,\, $i=2,3,...,N-1$
serves to control the extent of effective anharmonicity in the system and it can be interpreted as 
the intramolecular vibrational relaxation rate constant.
In what follows we set $\gamma_I=\gamma_{i}$ for the internal baths.
When $\gamma_{I}$ is large, phonon-phonon scattering effects are strong, while
the model reduces to the harmonic limit when $\gamma_{I}=0$.
The dissipation and the fluctuating force satisfy the relation ($k_B=1$),
\bea
&&\frac{1}{2}\langle\eta_s(\omega)\eta_m(\omega')+\eta_s(\omega')\eta_m(\omega)\rangle
\nonumber\\
&&=\frac{\hbar\omega }{2\pi}\,\Gamma_s(\omega) \,\textrm{coth}\bigg(\frac{\hbar\omega}{2T_s}\bigg) \, \delta(\omega+\omega')\, \delta_{s,m}.
\label{eq:noise}
\eea
The steady-state thermal current  at each bath is obtained from the correlation function 
$\langle K_{SB}X_B \dot{X}^T_S\rangle$ \cite{Roy06,DharRev}, which accounts for the input power.
In the language of Green's functions, 
the phonon heat current towards the $s$ bead, from the attached inner bath is given by \cite{DharSaito}
\begin{widetext}
\bea
F_s=\sum_{m=1}^{N}\int_{-\infty}^{\infty}d\omega\,\Gamma_s(\omega) \, \Gamma_m(\omega)\, \lvert[G(\omega)_{s,m}]\rvert^2  \, \frac{\hbar\omega^3}{\pi}\, [f(\omega, T_s)-f(\omega, T_m)].
\label{eq:current}
\eea
\end{widetext}
This is a multi-terminal transport expression. Incoming energy from the $s$ bath can transfer to any of the other terminals.
Below we refer to this expression as the `QSCR result'. 
The Green's function $G(\omega)$ is the inverse of a tridiagonal matrix with off-diagonal elements 
$-\omega_0^2$ and diagonal elements $2\omega_0^2-\omega^2-i\omega\Gamma(\omega)$.
$f(\omega,T)=[e^{\hbar\omega/T}-1]^{-1}$ is the Bose-Einstein distribution function at temperature $T$.
For simplicity, we set the atomic masses at $m_s=1$.
The essence of the QSCR method is the condition of null energy leakage from the system
towards each of the SC baths,
\bea
F_i=0, \qquad i=2,3,\,...\,N-1.
\label{eq:SCcondition}
\eea
This condition enforces energy conservation: energy leaving the $L$ solid reaches the 
$R$ contact. Meanwhile, the inner baths mimic phonon scatterings in the molecule
since the phonon number is not conserved in this transport process.
Eq. (\ref{eq:SCcondition}) is  nonlinear in the inner baths' temperatures, and thus 
a closed-form analytical solution for these temperatures is generally inaccessible. 
The roots of this equation, which are the temperatures of the SC baths,
can be solved for numerically.
Plugging these temperatures back into $F_1$
(or equivalently, into $F_N$), yields the steady-state net heat current across the junction, $J=F_1=-F_N$.

The advantage of the QSCR method lies in its applicability across different regimes: 
\\
(i) From far from equilibrium to linear response. 
At the nanoscale, large temperature drops may develop over a short gap.
A brute force solution of Eq. (\ref{eq:SCcondition}) 
provides the net heat current in general nonequilibrium settings.
However, the computational problem can be greatly simplified in
the linear response regime, $(T_L-T_R)/ T_a\ll1$, where  $T_a\equiv(T_L+T_R)/2$.
When the temperature drop along the chain is small, we can Taylor expand 
$(f_s-f_m) \sim \partial f/\partial T_a \times (T_s-T_m)$.  
Eq. (\ref{eq:current}) then reduces to what we
refer to as the Quantum Linear-Response (QLR) expression,
\bea
F_s^{QLR}&=&\sum_{m=1}^{N}\int_{-\infty}^{\infty}d\omega\,\Gamma_s(\omega) \, \Gamma_m(\omega)\, \lvert[G(\omega)_{s,m}]\rvert^2  \, \frac{\hbar\omega^3}{\pi}
\nonumber\\
&\times& \frac{\hbar\omega}{4T^2_a}\,\textrm{csch}^2\bigg(\frac{\hbar \omega}{2T_a}\bigg)\,(T_s-T_m).
\label{eq:QLRcurrent}
\eea
Here, ${\rm csch} (x)=\frac{1}{{\rm sinh } (x)}$ is the Hyperbolic cosecant function.
\\
(ii) From the quantum regime to classical behavior.
The chains' frequencies may be high relative to the averaged temperature, requiring a quantum treatment.
In contrast, in the classical (C) limit,
 $f(\omega, T) \sim T/(\hbar\omega)$, thus simplifying  Eq. (\ref{eq:current}) to 
\bea
F_s^{C}=\sum_{m=1}^{N}\int_{-\infty}^{\infty}d\omega\,\Gamma_s(\omega) \, \Gamma_m(\omega)\, \lvert[G(\omega)_{s,m}]\rvert^2  \, \frac{\omega^2}{\pi}\,(T_s-T_m).
\nonumber\\
\label{eq:Classicalcurrent}
\eea
Equations (\ref{eq:QLRcurrent}) and (\ref{eq:Classicalcurrent}) 
are linear in the SC baths temperatures. 
We rearrange these equations as $F_s=\sum_{m}C_{s, m}(T_s-T_m)$, where $C_{s, m}$ 
includes the frequency integration. 
The solution to the algebraic equations $F_s=0,\,s=2, ..., N-1$ 
is formally given by \cite{Segal09}
$T=A^{-1}\,\nu$,
where, $A$ is an $(N-2)\times(N-2)$ matrix with diagonal elements 
$\sum_{m \neq s}C_{s, m}$ and nondiagonal elements $-C_{s, m}$. 
The vector $\nu$ is defined as $\nu_s=C_{s, 1}T_1+C_{s, N}T_N$, 
and the vector $T$ contains the steady-state temperatures at 
the  inner baths. 
From the vector $T$, the net heat current $F_1$ can be readily computed.
\\
(iii)  From harmonic to anharmonic chains / from ballistic to diffusive transport mechanisms.
The phenomenology of anharmonic interatomic potentials 
is introduced in the QSCR method through the SC reservoirs.
Suppressing $\gamma_I$, heat transport across the system
becomes increasingly dominated by the elastic-harmonic (H) contribution,
\bea
F_1^{H}&=& \int_{-\infty}^{\infty}d\omega\,\Gamma_1(\omega) \, \Gamma_N(\omega)\, \lvert[G(\omega)_{1,N}]\rvert^2  \, \frac{\hbar\omega^3}{\pi}\,
\nonumber\\
&\times& [f(\omega, T_1)-f(\omega, T_N)].
\label{eq:currentH}
\eea
We recover the Landauer formula for heat transport when $\gamma_I=0$. In Appendix A we organize Eqs. (\ref{eq:current})
and (\ref{eq:currentH})  in the language of the transmission function.
The QSCR method interpolates between the coherent ballistic limit of energy transport and the diffusive
limit by tuning the inelastic rate constant $\gamma_I$ \cite{Bonetto04,DharRev,Roy08}.
\begin{figure}[htbp]
\hspace{-14mm}
\includegraphics[width=9.5cm]{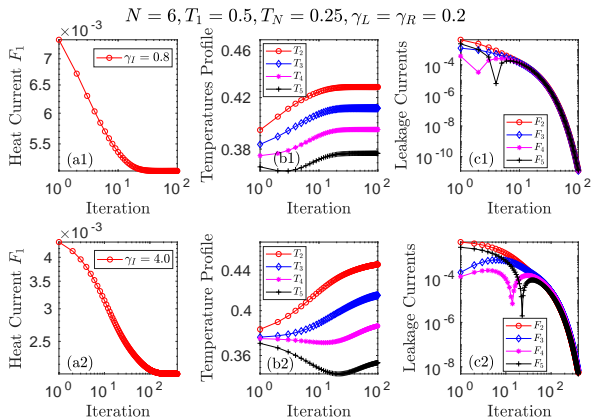} 
\raggedleft
\caption{Convergence of the QSCR method with increasing number of iterations 
using structureless baths: (a) net heat current, 
(b) inner temperature profile, 
and (c) leakage currents.
Parameters are $\gamma_I=0.8$ (top)  and  $\gamma_I=4.0$ (bottom).}
\label{Fig2}
\end{figure}

\begin{figure}[!htbp]
\hspace{-14mm}
\includegraphics[width=9.5cm]{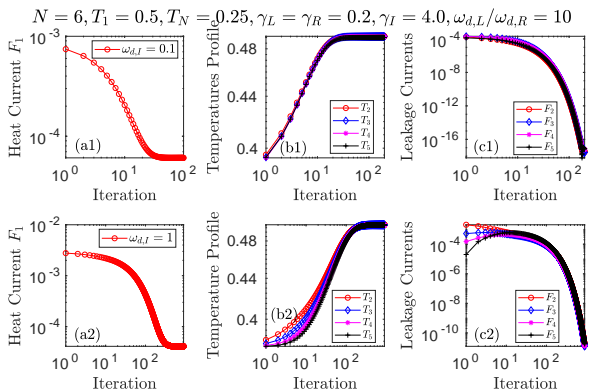} 
\caption{Convergence of the QSCR technique with increasing number of iterations using structured baths: 
(a) net heat current, (b)  temperature profile,  and (c) leakage currents.
Parameters are $\omega_{d,L}=1$, $\omega_{d,R}=0.1$ with
$\omega_{d,I}=0.1$ (top) and $\omega_{d,I}=1$ (bottom). 
}
\label{Fig3}
\end{figure}

\section{Numerical procedure and convergence tests} 

The numerical scheme for solving the QSCR equations was 
discussed in Ref. \cite{Malay11}. 
We review the procedure here. 
We solve Eq. (\ref{eq:SCcondition}) using the Newton-Raphson technique \cite{numR}.
Given a well-behaved function $f(x)$, the root $r$ satisfying
$f(r)=0$ can be obtained iteratively from an initial guess $x_0$, 
$x_{n+1}=x_n-\frac{f(x_n)}{f'(x_n)}$, where $f'(x_n)$ is the first derivative of $f(x)$ at $x_n$.
The procedure is generalized to solve Eq. (\ref{eq:SCcondition}), which is an $N-2$ 
nonlinear system \cite{Malay11}. 
For the initial guess we choose the average temperature, 
 $T^{(0)}_{i}=(T_1+T_N)/2, \,\, i=2, .., N-1$. 
At each iteration $k$, the temperatures of the SC reservoirs are updated according to
\bea
T^{(k+1)}_{i}=T^{(k)}_{i}-\sum_{j=2}^{N-1}(D^{-1})_{i,j}F_j(T^{(k)}).
\label{eq:MatrixNR}
\eea
Here, $T^{(k)} $ is the vector of temperature profile after the $k$th iteration. 
$D$ is the Jacobian matrix with the elements $D_{i,j}={\partial F_i}/{\partial T_j}$.

We verify convergence by examining three quantities as we iterate: 
(i) The temperature profile of the SC reservoirs should approach a fixed value between $T_1$ and $T_N$. 
Agreeably, (ii) the net heat current across the system, $|F_1|=|F_N|$, should focalize to a finite value.
(iii) The individual leakage currents, from the SC baths, should become 
vanishingly small relative to the net heat current,
$|F_{i=2, ...,N-1}|\ll|F_1|$. 
Convergence is declared when the relative errors, for 
the net heat flow and the temperature profile, are smaller than $10^{-6}$,
$Error(x_n)=(x_n-x_{n-1})/x_n$.

The number of iterations required to achieve convergence
depends on the temperature difference, chain length, 
averaged temperature, phonon scattering rate $\gamma_I$, and the cutoff frequency $\omega_d$.
For example, as $\gamma_I$ increases, the SC baths become more impactful in scattering phonons and 
convergence becomes more challenging. 
This point is illustrated in Fig. \ref{Fig2}, where we study a chain with $N=6$ particles
connected to structureless  reservoirs  (no cutoff).  We converge
after $50$ iterations when  $\gamma_I\sim \gamma_{L,R}$.
However, convergence is reached only after as many as $200$ iterations when  $\gamma_I>\gamma_{L,R}$.
It is interesting to note that convergence is not necessarily monotonic for all sites. 

Convergence is facile if the cutoff frequency of the SC reservoirs is rather low,
since predominantly, only modes below the cutoff frequency suffer from scattering effects, while modes above it
are left intact.
As a representative example to simulations presented in Sec. \ref{sec-res2},
in Figure \ref{Fig3} we study a chain with $N=6$ particles 
connected to two thermal reservoirs with different cutoff frequencies. 
The mismatch is significant, 
and the ratio of the Debye frequencies of the hot to cold reservoir is  
$\omega_{d,L}/\omega_{d,R}=10$. 
In addition, we set a finite value for $\omega_{d,I}$ for all the internal baths.
We find that when $\omega_{d,I}<\omega_0$ (top panels), the convergence is fast
relative to the case of $\omega_{d,I}\sim \omega_0$ (bottom panels). 
We further tested (not shown) the role of the temperature bias on convergence.
As expected, when the hot and cold baths temperatures largely differ, 
more iterations are required to converge. 
In all our simulations we set the masses to 1, and use $\omega_0=1$.

\begin{figure*}[ht]
\centering
\begin{subfigure}{.5\textwidth}
\centering
\includegraphics[width=8.0cm]{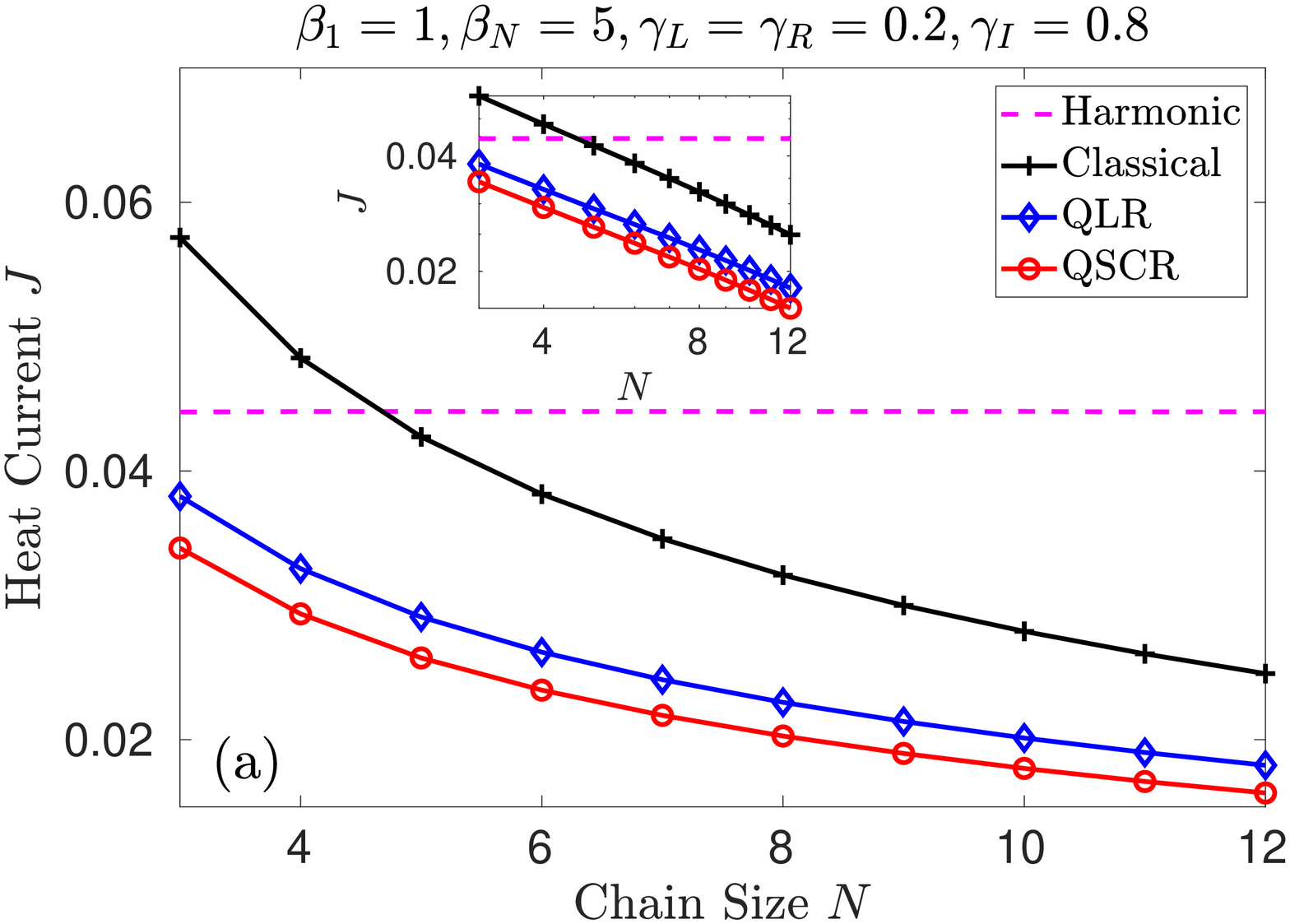} 
\end{subfigure}%
\begin{subfigure}{.5\textwidth}
\centering
\includegraphics[width=8.0cm]{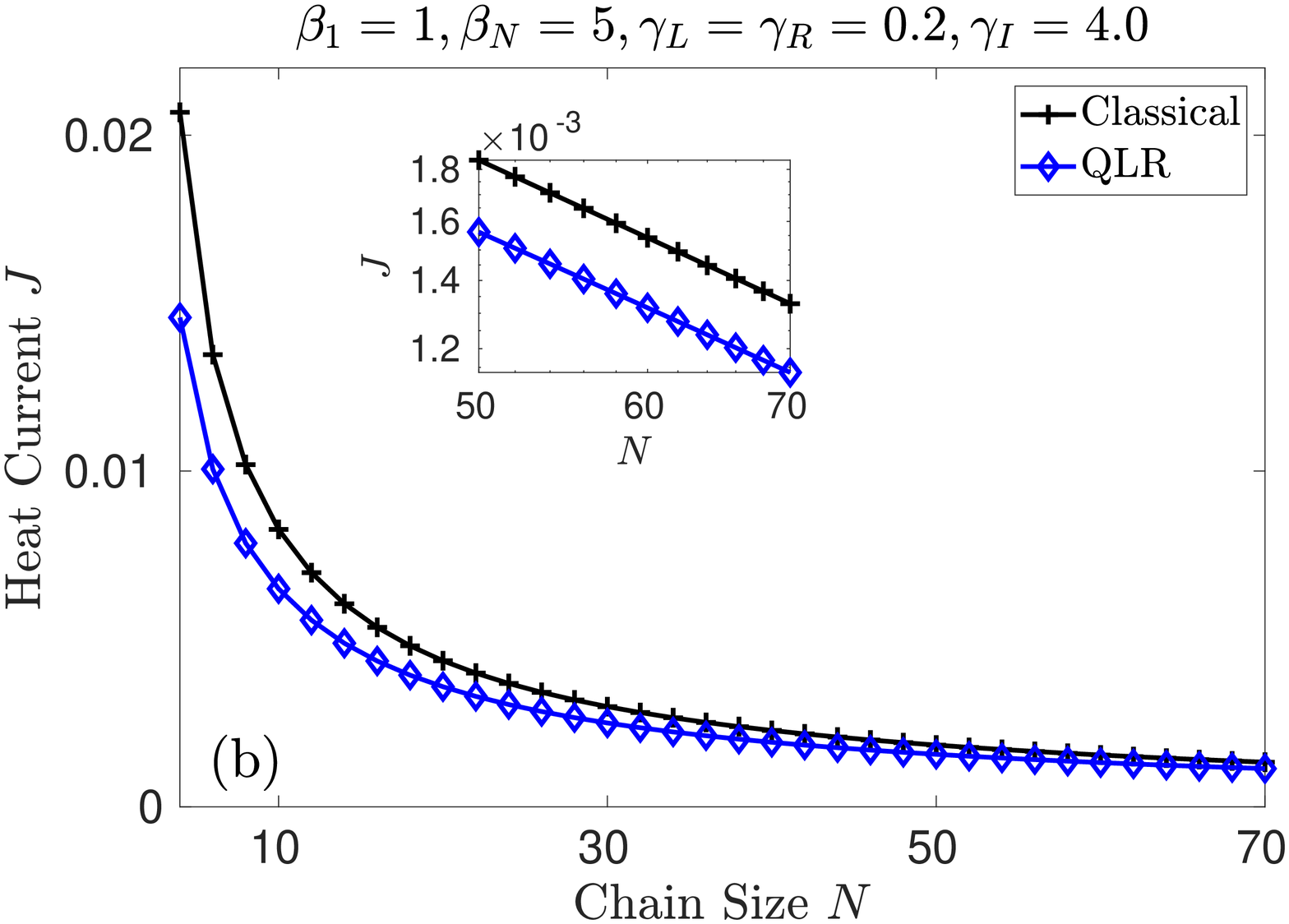}  
\label{fig:sub2}
\end{subfigure}
\caption{Heat current as a function of chain length at low temperatures 
using the QSCR method (circle),  quantum linear-response  (diamond), classical (+), and harmonic (dashed) calculations for 
(a) short chains with weak $\gamma_I$  and (b) long chains with strong $\gamma_I$. The insets display the corresponding log-log scale plot.}
\label{Fig4}
\end{figure*}

\begin{figure*}
	\begin{subfigure}{.5\textwidth}
		\centering
		\includegraphics[width=8.0cm]{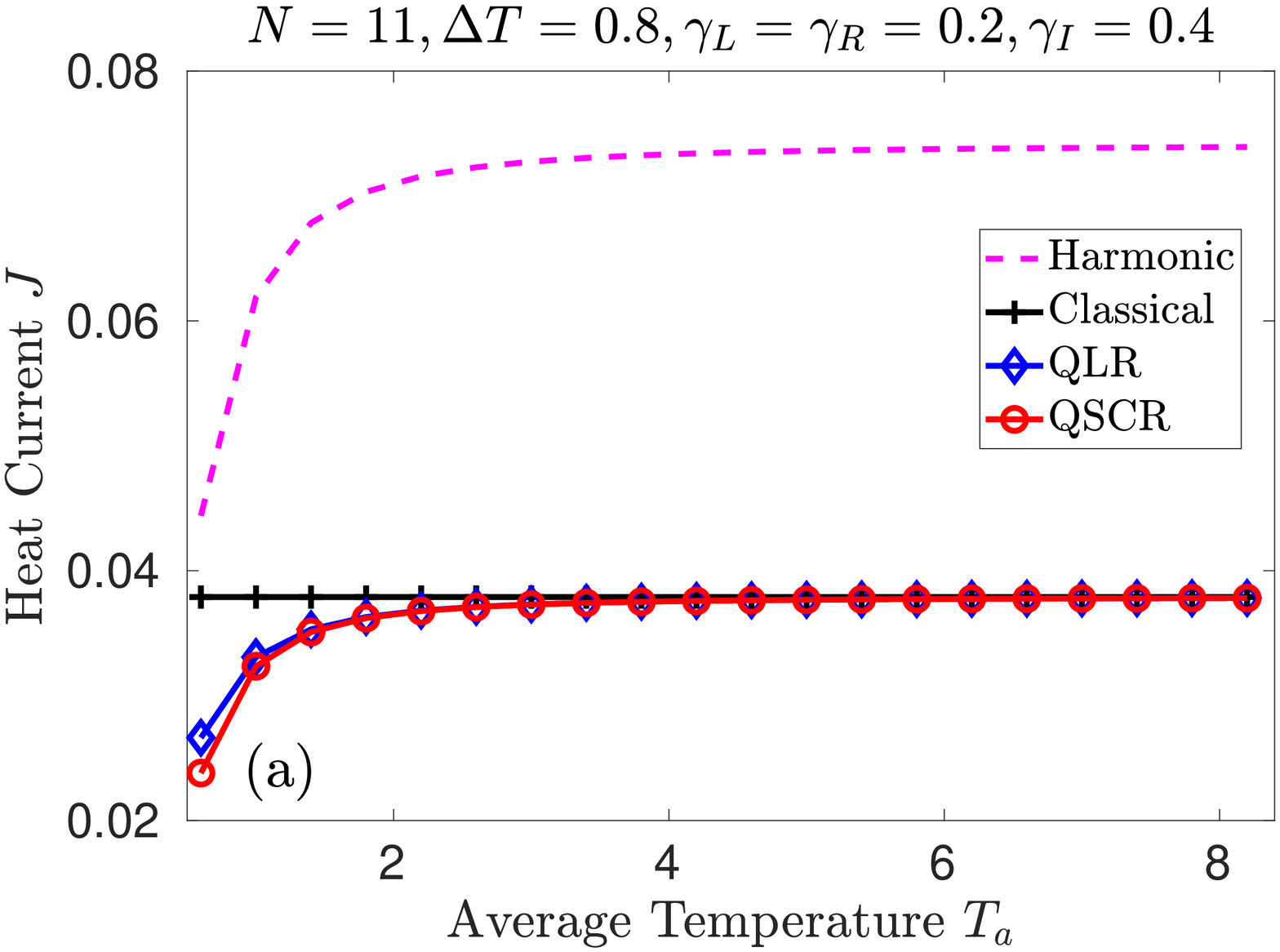} 
	\end{subfigure}%
	\begin{subfigure}{.5\textwidth}
		\centering
		\includegraphics[width=8.0cm]{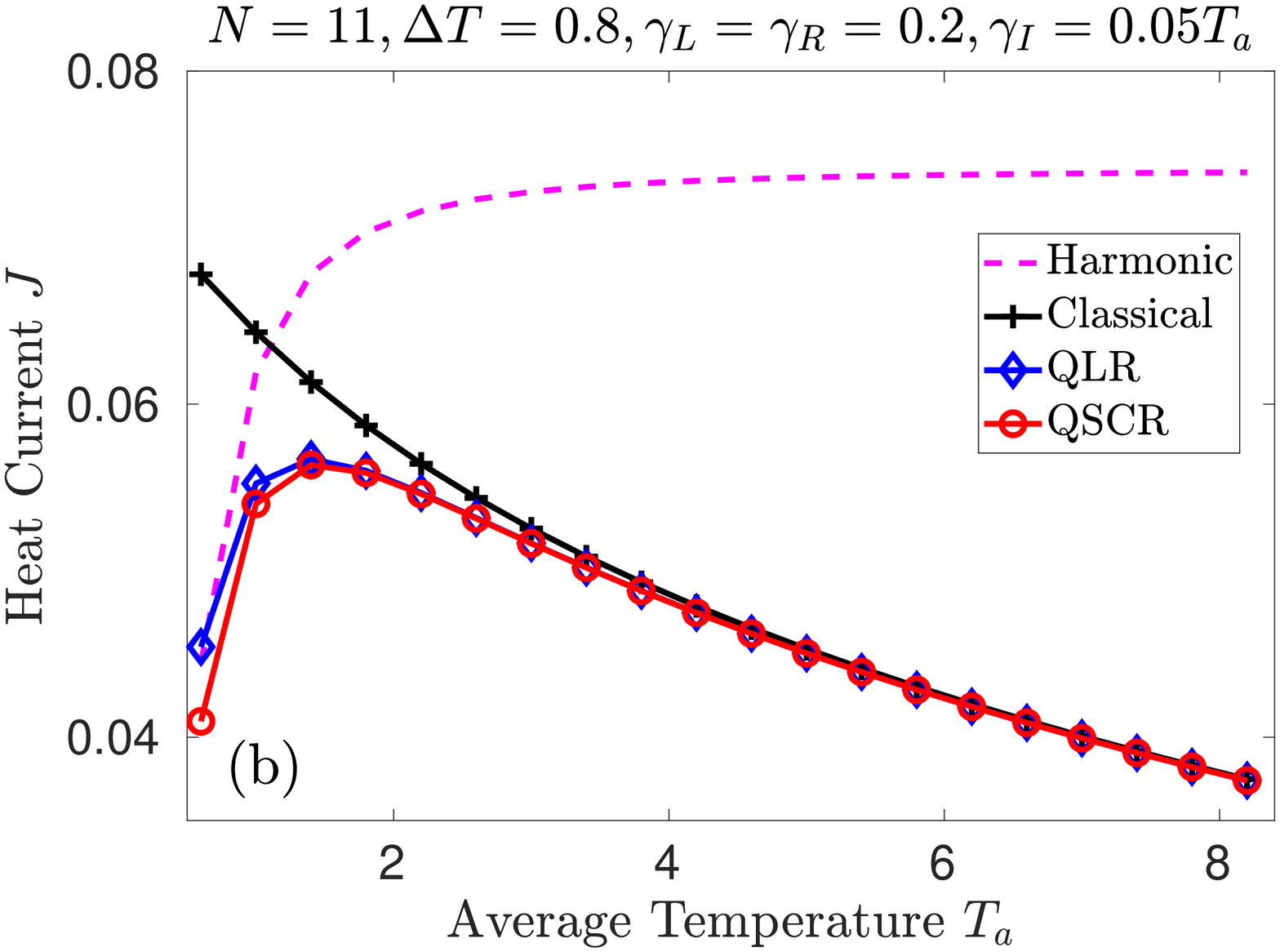}   
\end{subfigure}
\caption{ Heat current as a function of averaged temperature 
using the QSCR method (circle), quantum linear-response (diamond), classical (+), and harmonic (dashed)
limits. (a) Temperature-independent vs. (b) temperature-dependent phonon scattering rates.}
\label{Fig5}
\end{figure*}

\begin{figure*}
\begin{subfigure}{.5\textwidth}
\includegraphics[width=8.0cm]{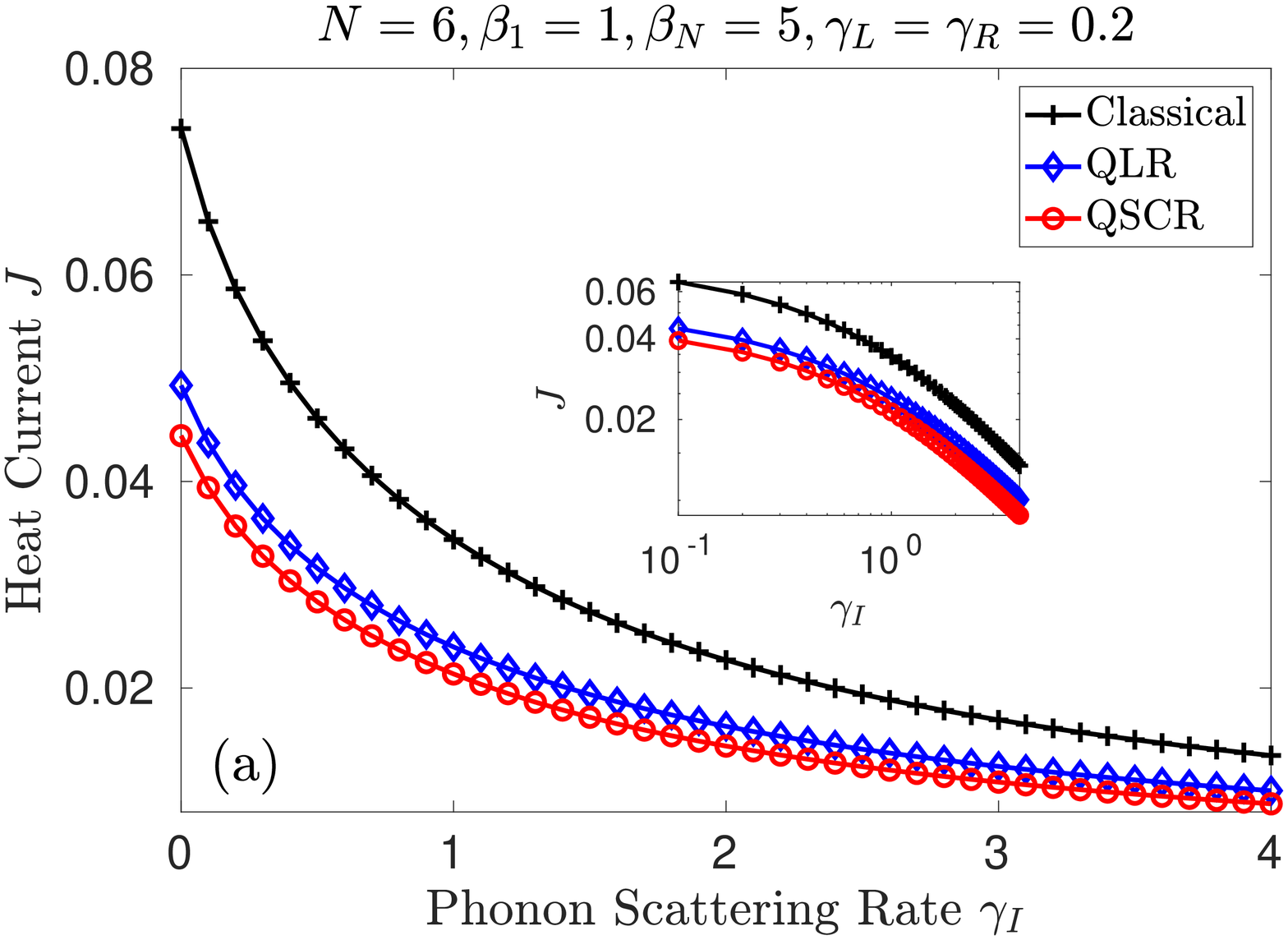} 
\end{subfigure}%
\begin{subfigure}{.5\textwidth}
\includegraphics[width=8.0cm]{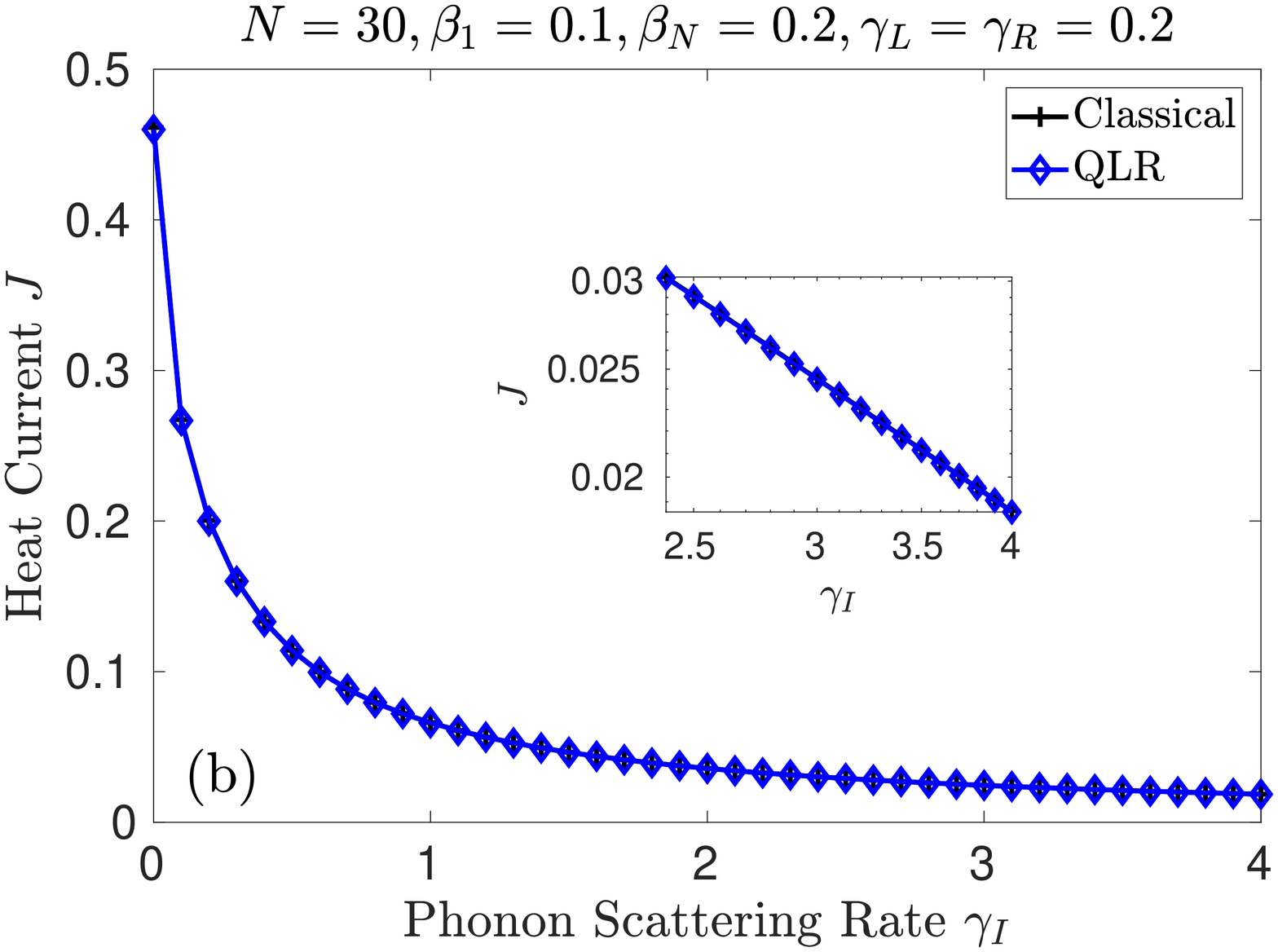} 
\end{subfigure}
\caption{ Heat current as a function of phonon-phonon scattering  rate $\gamma_I$
using the QSCR (circle), QLR (diamond), and classical (+) calculations.
We study (a) a short chain at low temperature, and (b) a long chain at high temperature.
The insets display the corresponding log-log scale plot. }
\label{Fig7}
\end{figure*}

\section{Validation of the QSCR method}
\label{sec-res1}

In this Section we study the behavior of the heat current as a function of molecular length, temperature, and
phonon-phonon scattering rate $\gamma_I$. We illustrate that the QSCR method  captures qualitatively-correctly
the heat transport behavior across anharmonic MJs.
While we do not have a benchmark, first-principle simulation of quantum thermal conduction in anharmonic MJs,
studies of heat transport in nanostructures \cite{CahillRev1,CahillRev2,Gang}, 
such as self-assembled monolayers, amorphous polymers and nanowires,
display characteristic trends, including: (i) crossover from ballistic to diffusive dynamics 
and the development of the Fourier's law of thermal conduction with increasing length,
(ii)  turnover behavior of the heat current with temperature  \cite{Kim}.

We compare results from the following calculations:
(i) QSCR method, referring to the numerical solution of Eqs. (\ref{eq:current})- (\ref{eq:SCcondition}). 
This nonlinear method treats far from equilibrium cases (large $\Delta T$), effective anharmonicity
(large $\gamma_I$), and quantum effects ($\hbar \omega_0/T_a>1$, introduced through the statistics of the Bose-Einstein distribution function).
(ii) Quantum linear-response approximation, Eq. (\ref{eq:QLRcurrent}), which is valid when only a small temperature gradient develops.
(iii) Classical limit, $T_a>\hbar \omega_0$, using Eq. (\ref{eq:Classicalcurrent}).
 (iv) Quantum harmonic case, Eq. (\ref{eq:currentH}), valid when $\gamma_I=0$.

\subsection{Length dependence}

The behavior of the heat current with chain length is displayed in Fig.  \ref{Fig4}(a).
For short molecules, the conductance of quantum harmonic systems is generally non-monotonic 
with length \cite{SegalRev, Segal03,Pauli}, 
and it depends on the normal mode spectrum, temperature, the nature of the contacts and the properties of the solids.
Here, given the simplicity of the model (ordered 1D chain and $\omega_d\to \infty$), we observe a ballistic behavior 
with $J\propto N^0$ for harmonic systems.
The effect of anharmonicity, as captured in the QSCR method, is significant in this respect: It leads to the decay of
the current with molecular length, reflecting an increase in the thermal resistance.
Another observation from Fig.  \ref{Fig4}(a)  is that classical simulations overestimate the current.
In the classical regime the phonon distribution is proportional to the temperature 
$(f\rightarrow \frac{ T}{\hbar \omega})$, therefore all phonon modes become active \cite{Majumdar15,Amon08}. 
The quantum linear-response calculation is not accurate, as expected, but it is qualitatively close to the QSCR calculations.

We  examine whether the SC reservoirs can emulate Fourier's law of thermal conduction, 
that is, $J \propto \frac{\Delta T}{N}$ \cite{DharRev}.
The chains in Figure \ref{Fig4}(a) are rather short, and as a result the temperature drop along the chain is relatively large and we deviate from 
the diffusive scaling;
a log-log plot of conductance-length (see inset) reveals the slope $\sim -0.5$.
Longer molecules are studied in Figure \ref{Fig4}(b), where we find that (inset) $J\propto N^{-0.95}$, approaching
the Fourier's law behavior. Simulations here are done using the QLR expression since converging the full QSCR method
 is rather costly when $N$ is large.

\subsection{Temperature dependence}

The behavior of the thermal current with temperature is displayed in Figure \ref{Fig5}. We bias the junction
following $T_{L,R}=T_a\pm\Delta T/2$.
The low-temperature trend is expected: Classical predictions overestimate quantum simulations.
QLR calculations slightly exceed the heat current, as calculated by the QSCR method, since the thermal bias is relatively large at low temperatures. Harmonic simulations upper bound the current \cite{Ed}. 
Most notably, at high temperatures, similarly to harmonic predictions, the QSCR results saturate.
In contrast, bulk \cite{Ashcroft} and nanostructure materials \cite{Mingo} 
typically display a turnover behavior with temperature:  
The thermal conductance increases with $T_a$ to a certain power at low temperature, explained 
by the increase in specific heat. At high temperatures (beyond $T_a\sim \hbar\omega_0$), the conductance decays 
with $T_a$ due to the growing importance of (inelastic) Umklapp phonon-phonon scattering process.
Similarly, we expect that in MJs anharmonicity should become more impactful 
as the temperature is increased: As more modes become active, more pathways for 
redistributing the vibrational energy open up (phonons splitting, recombination), 
resulting in the enhancement of thermal resistance.

This turnover behavior can be realized in the QSCR method by making the internal scattering rates
a function of temperature, $\gamma_I(T_a)\propto T_a^p$, $p>0$.
With that, the potency of anharmonic effects grows as we increase the temperature, eventually suppressing the current.
This behavior is exemplified in Fig. \ref{Fig5}(b) using $\gamma_I\propto T_a$.
Since our system is low-dimensional, we roughly expect $J \propto T_a$ at low temperature, 
and $J \propto T_a^{-n}$ with $ 1<n<2$ at high temperature \cite{Ashcroft}. 
In Appendix B we show that by modifying the functional form of $\gamma_I(T_a)$, we change the current-temperature scaling.
Overall,  increasing $\gamma_I$ (essentially, the IVR rate constant) with temperature leads to a turnover behavior.
The specifics of the scaling functions and the transition behavior derive from  the functional form of $\gamma_I(T_a)$.

\subsection{Inelastic phonon-phonon scatterings}

The scattering rates of the SC baths $\gamma_I$ control the extent of anharmonicity in the model.  
As we increase this parameter we effectively enhance phonon splitting and recombination processes.
As well, instead of the direct harmonic left-to-right phonon transmission dynamics, 
excitations are scattered between the SC baths, thus the thermal resistance grows with $\gamma_I$.
This trend is observed in Fig. \ref{Fig7}(a).
We again note that, similarly to Fig. \ref{Fig4}(a), the heat current is overestimated by the classical limit,
and it is slightly overvalued using the QLR method. 
Dhar \textit{et. al.} \cite{Roy06}, following Ref. \cite{Bonetto04} showed that in the classical thermodynamic 
limit, the thermal conductance $K_T=J/\Delta T$ should be proportional 
to the inverse of the phonon scattering rate, $K_T\propto\gamma_I^{-1}$. 
A log-log  plot shown in the inset of Figure \ref{Fig7}(a) displays deviation from this trend,
indicating that the current is still largely influenced by the direct (harmonic) left-to-right transmission
contribution.
In contrast, as we show in Figure \ref{Fig7}(b), in long chains at high temperature  
 $J \propto \gamma_I^{-0.95}$, approaching the theoretical limit \cite{DharRev, Bonetto04}. 


\subsection{Physical units: alkane chains}

At this point, it is useful to express our results in physical units.
We note that the memory function $\Gamma(\omega)$,
and therefore the coefficients $\gamma_s$ have the dimension of frequency. 
With that, we readily verify that the heat current in Eq. (\ref{eq:current}) 
takes the dimension of energy over time, as required.
In simulations, we set $\hbar \omega_0=1$. 

To simulate an alkane chain, we consider a mode of frequency 1000 cm$^{-1}$, which is representative for the single bond
carbon-carbon stretching motion. With this frequency  ($\omega_0=3\times$10$^{13}$ Hz), 
we  focus for example on Fig. \ref{Fig4}(a).
The parameters  $T_a\sim 1$,  $\gamma_{L,R}=0.2$, $\gamma_{I}=0.8$, 
and $\Delta T=0.8$ result in the heat current $J\sim0.04-0.02$ for a chain with $N=4-12$ particles.
In physical units, based on $\omega_0=3\times$10$^{13}$ Hz, these numbers translate to
$T_a$= 230 K, $\Delta T=180$ K,  $\hbar \gamma_{L,R} =4$ meV for the molecule-surface contact energy, 
$\hbar \gamma_{I}=16$ meV for the vibrational energy scattering rate, 
or 25 ps$^{-1}$.
The heat current corresponding to $J=0.03$ translates to $J=3 \times 10^{-9}$ W, resulting in the thermal 
conductance  $K_T=16$ pW/K, where $K_T=J/\Delta T$.
This number is comparable to values received in experiments on alkane chains with 2-18 methylene units, 
reporting $K_T\sim 5-25$ pW/K \cite{Gotsmann,Majumdar15}.
Note that the value used here for the 
inelastic rate constant $\gamma_I$ exceeds estimates for alkane chains
by about an order of magnitude  \cite{Leitner16}. 
We amplify this parameter here so as to observe signatures of inelastic effects in short-periodic molecules. 
When using a smaller value for $\gamma_I$, as more appropriate for alkanes, the QSCR method produces result that concur with the harmonic limit.

In the next section, we furthermore consider Debye solids with characteristic frequencies 
in the range of 0.1-1, in units of $\omega_0$, translating to $\omega_d=3\times$ 10$^{12}$ - 3$\times$10$^{13}$ Hz. In the language of Debye temperature these numbers correspond to  $T_D =22 - 220 $ K.
The metals used in Ref. \cite{Majumdar15} are
gold, silver,  platinum are palladium, with Debye temperatures 
$T_{{D}_{\rm Au}}=170$ K,  $T_{{D}_{\rm Ag}}= 215$,
$T_{{D}_{\rm Pt}}=240$ K, and  $T_{{D}_{\rm Pd}}= 275$ K  \cite{Kittel}.
Our   
calculations in Sec. \ref{sec-res2} are therefore relevant 
to the phonon-mismatch thermal conductance
experiment reported in Ref. \cite{Majumdar15}.
We emphasize though that the main goal 
of this paper is to present a viable technique for simulating
quantum heat transport in molecular junctions, rather than to reproduce a specific experiment.

\begin{figure}[ht]
\includegraphics[width=9cm]{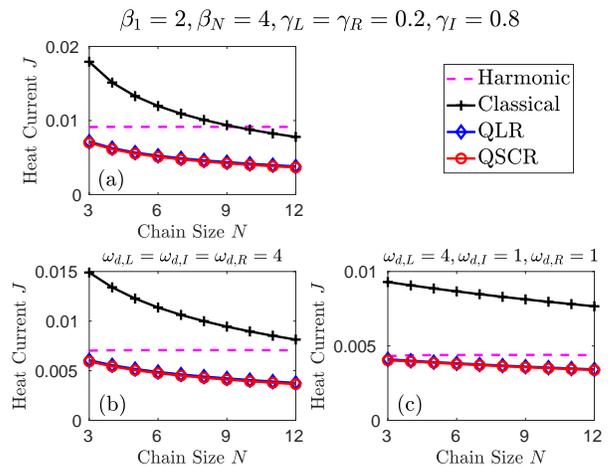}  
\caption{Heat current as a function of chain length with (a) structureless reservoirs,
(b) Debye baths without mismatch, and (c) mismatched Debye reservoirs.
In all panels we display the QSCR (circle), quantum linear-response (diamond), classical (+) and
harmonic (dashed) calculations.}
\label{Fig8}
\end{figure}
\begin{figure}[htbp]
\includegraphics[width=8.0cm]{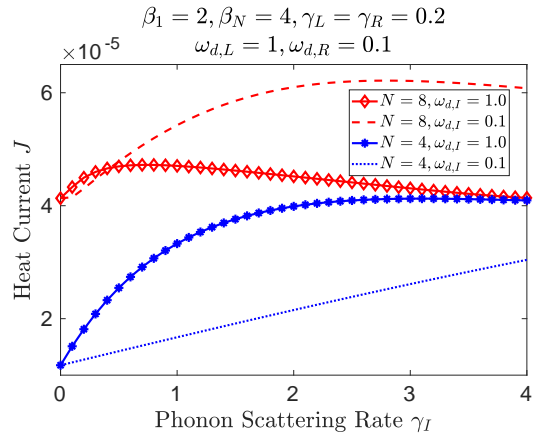} 
\caption{Heat current as a function of phonon scattering rate at low temperatures for $N=4,8$.
The mismatch between the phonon spectra of the hot and cold baths is high, $\omega_{d,L}/\omega_{d,R}=10$.}
\label{Fig9}
\end{figure}

\section{Thermal transport between solids with a phonon mismatch}
\label{sec-res2}

In the previous section we demonstrated that the QSCR method is suitable to describe anharmonic effects in quantum thermal transport, but so far we used structureless baths.
In this section, we study heat transport between surfaces with structured, possibly distinct phonon spectra.
Our modelling is simple: We use the Debye model in which the phonon spectra is characterized by a
soft cutoff frequency $\omega_d$,
Eq. (\ref{eq:friction}). This simple setup in fact is compound as transport involves anharmonicity, 
quantum effects, mismatched phonon spectra and a far from equilibrium situation. 
Particularly, we simulate below the heat current as a function of the phonon mismatch between reservoirs.
While in harmonic systems the current should drop when raising the mismatch,
inelastic scatterings can assist in overcoming it. 
Nevertheless, the prediction of classical molecular dynamics simulations,
 of an enhanced thermal conductance with the increase in phonon 
mismatch are surprising, and in conflict with measurements \cite{Majumdar15}.

In Figure \ref{Fig8} we study three cases: (a) structureless baths, as used in Sec. \ref{sec-res1}, 
(b) Debye baths with identical Debye frequencies, and (c) Debye baths with a vibrational mismatch, 
$\omega_{d,L}\neq \omega_{d,R}$. We find that as we progress from (a) to (b) to (c), 
the heat current as predicted by the QSCR method is reduced.
 In particular, in Fig. \ref{Fig8}(c)  the current is essentially dominated by the harmonic contribution. 
 In this case, high frequency modes from the left hot bath cannot cross to the right cold side---unless energy
 is redistributed in the  inner baths. Since this process is not very efficient for short molecules, 
 vibrational mismatch leads to a significant suppression of current.

The dual role of the SC reservoirs (thus anharmonicity) is illustrated in Fig. \ref{Fig9}: 
On the one hand, phonon scattering events increase the resistance thus reduce the current, but on the other hand,
inelastic scatterings can assist to overcome a vibrational mismatch. 
Elaborating: Let us recall first what we observe in Fig. \ref{Fig7},  that the heat current 
decays as we increase $\gamma_I$---for structureless baths.
As we show in Fig. \ref{Fig9}, the situation is different when the solids have mismatched phonon spectra.
First, when $\gamma_I$ is small the current increases with $\gamma_I$ reaching values
 beyond the harmonic limit ($\gamma_I=0$).
In this regime, the SC baths act to overcome the mismatch, 
and anharmonic effects are beneficial for transport.
However, beyond a certain value of $\gamma_I$, which naturally depends on molecular length and temperature,
the main effect of the SC baths is to dampen the current due to multiple scattering processes 
that overall enhance the resistance.
As expected, the crossover between the two regimes takes place at smaller $\gamma_I$ for longer molecules, 
see the comparison between $N$=4 and $N$=8 junctions.
This illustration, of the dual role of the SC baths, is one of the main results of the paper.

Next, motivated by the experiment reported in Ref. \cite{Majumdar15}, 
we consider MJs with dissimilar phonon spectra at the boundaries, 
captured by different Debye frequencies.
In what follows, we describe two such setups and study 
the influence of the vibrational mismatch on the heat conductance characteristics.


\subsection{Model I: Phonons up-conversion processes} 
\begin{figure}[htbp]
\includegraphics[width=9cm]{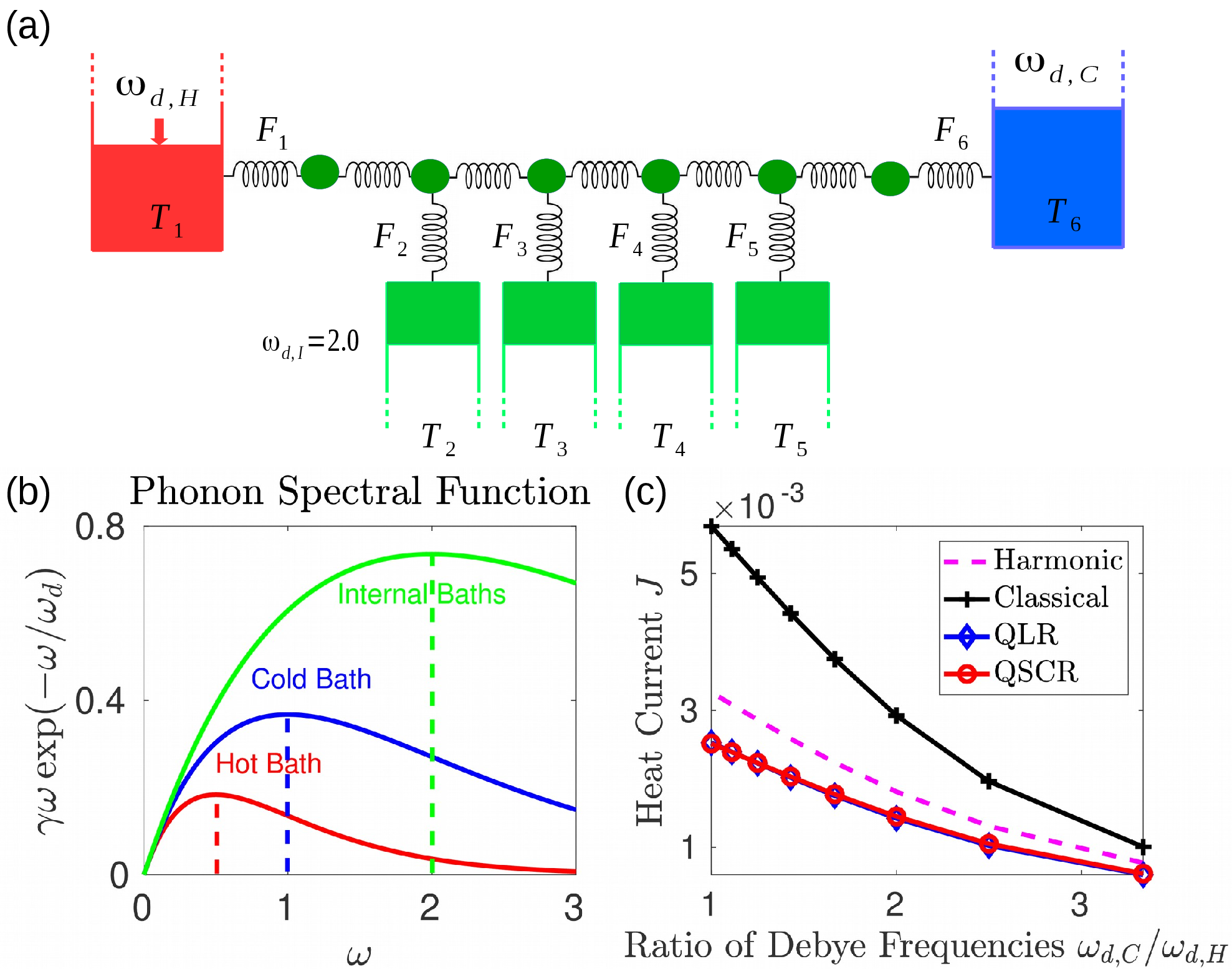} 
\caption{Model I for MJs with dissimilar phonon spectra.
(a) Scheme of the SC model of length $N=6$ with mismatched reservoirs,
 $\omega_{d,H}=1, 0.9,...,0.3$, and $\omega_{d,C}=1$, and hot and cold inverse temperatures,
 $\beta_1=2$, $\beta_N=4$, respectively.
(b) Example of the phonon spectral function with $\omega_{d,H}=0.5$,  $\omega_{d,C}=1$, and $\omega_{d,I}=2$.
(c) Heat current as a function of phonon mismatch.
We compare the QSCR (circle), QLR (diamond), classical (+) and harmonic (dashed) calculations.}
\label{Fig10}
\end{figure}

In the configuration of Fig. \ref{Fig10}(a) 
the cold bath is fixed with a Debye frequency of   $\omega_{d,C}=1$. For the hot baths different
solids are used; the Debye frequency of the hot bath is gradually reduced, starting from the same value as 
the cold bath.
The overlap between the phonon spectra of the reservoirs is exemplified in   Fig. \ref{Fig10}(b).
In this setup, phonons collisions to create a high frequency mode (up-conversion) are beneficial for transport.
Fig. \ref{Fig10}(c) shows that the heat current decreases as the vibrational mismatch 
(measured by the ratio of Debye frequencies) between the  
reservoirs increases, which is in agreement with experimental data \cite{Majumdar15}. 
Classical, QLR and harmonic calculations provide a similar trend. 
Furthermore, while the classical limit overestimates 
the heat current, its predictions are still in line with the QSCR method.

\subsection{Model II: Phonons down-conversion processes}

\begin{figure}[htbp]
\includegraphics[width=9cm]{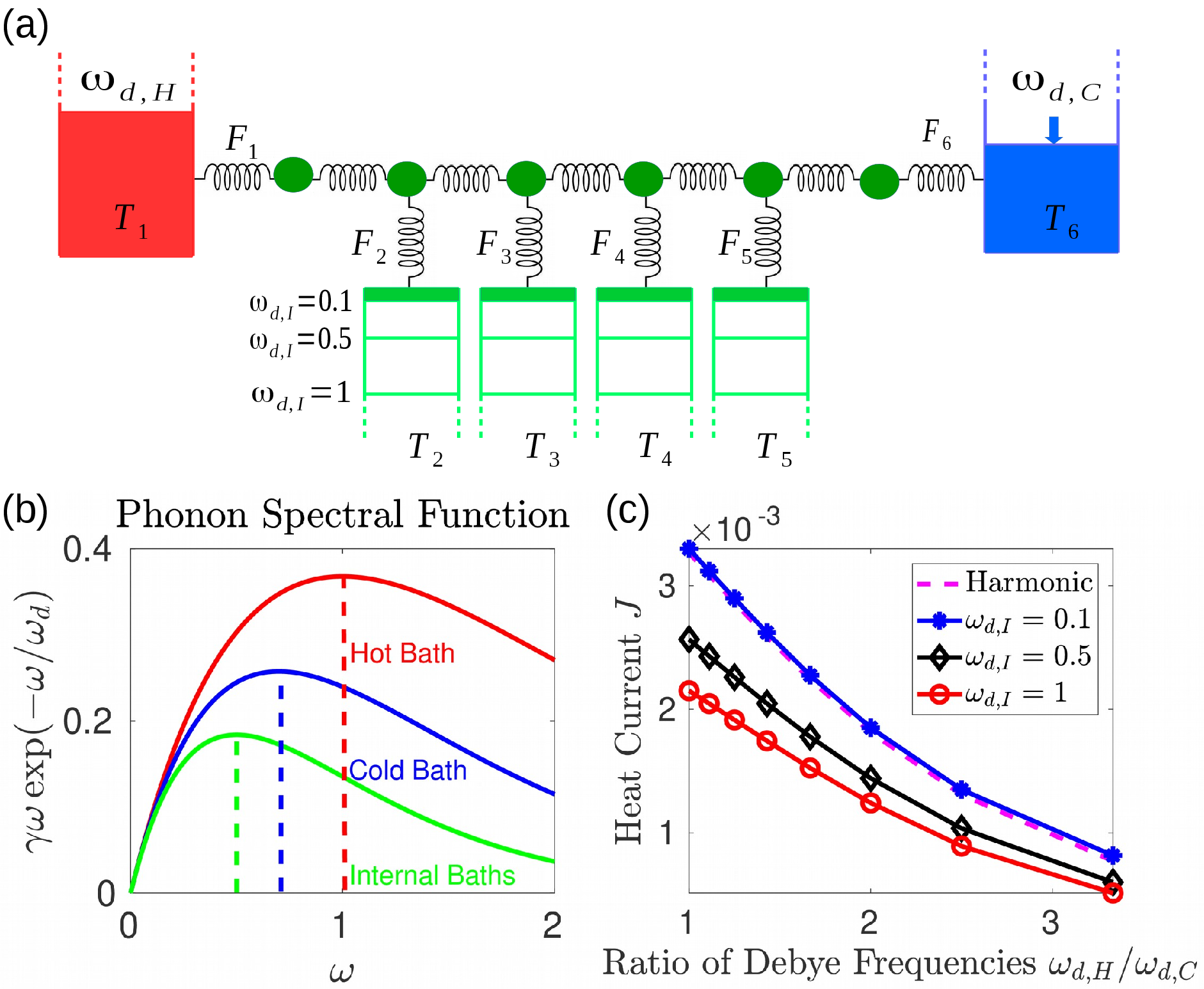} 
\caption{Model II for MJs with dissimilar phonon spectra.
(a) Scheme of an $N=6$  model with mismatched reservoirs,
$\omega_{d,H}=1$ and $\omega_{d,C}=1, 0.9,...,0.3$,
with inverse temperatures $\beta_1=2, \beta_N=4$. 
We vary the internal scatterings as $\omega_{d,I}=0.1, 0.5, 1$.
(b) Examples of phonon spectral function with  $\omega_{d,H}=1$, $\omega_{d,C}=0.7$ and  $\omega_{d,I}=0.5$.
(c) Heat current as a function of phonon mismatch.}
\label{Fig11}
\end{figure}

In the configuration of Figure \ref{Fig11}, the hot solid is fixed and its Debye frequency  is equal or higher than the cutoff 
 frequency of the cold bath. High frequency modes are therefore prohibited from crossing 
the system---in the harmonic limit.
Here, decay processes (phonons splitting to low frequency modes) are beneficial for transport.
Figure \ref{Fig11}(c) indicates that the heat current decreases as the vibrational mismatch between the solids increases. 
Furthermore, we examine the effect of the cutoff frequency of the SC baths on the heat current.
A low cutoff frequency for the SC baths impedes their ability to scatter phonons, thus results approach the
 harmonic limit as we show in Figure \ref{Fig11}(c). 
In contrast, when the SC baths accommodate the full spectra of modes ($\omega_{d,I}$ large), the thermal resistance increases and the heat current is reduced. Most notably, while the SC baths support down-conversion processes, their dominant effect is to dampen the current.

The two models confirm that the Landauer (harmonic) formalism captures reasonably well 
the transport behavior in the chain as a function of phononic mismatch for short molecules with $N=4-10$ particles, 
and that the current decays when increasing the phonon mismatch between the interfaces.
A similar conclusion was reached in Ref. \cite{Leitner16} 
using the Fermi Golden Rule to calculate phonon-phonon collision and decay processes.

It is worthwhile to mention that the effect of the solids' phononic spectral properties on the heat transport 
characteristics, for harmonic and anharmonic molecules, were analyzed in Ref. \cite{Yun10} using classical molecular dynamics simulations
with an explicit anharmonicity. Adopting spectral densities with certain high or low frequency phonon bands, it was found that the heat current
characteristics were highly sensitive to the solid-molecule vibrational frequency mismatch: 
While for harmonic molecules the current was fixed for long enough molecules $N>5$, 
for anharmonic systems it decayed (increased) with chain length for down (up) conversion processes.

\section{Summary}
 \label{sec-summ}

The simulation of quantum heat flow in anharmonic nanostructures is a challenging task. 
We demonstrated here the utility of the QSCR method, 
a treatment that emulates anharmonicity within SC reservoirs.
This tool reproduces the expected phenomenology of heat transfer in molecular junctions
as a function of length, temperature, and phonon scattering processes.
The QSCR method is simple to implement, and it can be applied to large-scale systems. 
This quantum-effective anharmonic method can be analyzed in the truly harmonic, classical,
and the linear response limits. Therefore, it allows one to asses the impact of these factors in thermal 
conduction, and smoothly interpolates
between the ballistic and diffusive transport regimes.

As an application, we studied the role of phononic mismatch on heat transfer in MJs, 
by using solids with different phonon spectra.
Anharmonicity as implemented by the SC baths assists in overcoming a vibrational mismatch, 
as we showed in Fig. \ref{Fig9}.
With increasing phonon mismatch, which is measured by the ratio of Debye
frequencies of the solids, 
the heat current decreased, in line
with experimental observations---and in contrast to classical 
molecular dynamics simulations \cite{Majumdar15}.
The suppression of the heat current with phonon mismatch as we observed in Figs. \ref{Fig10}-\ref{Fig11}
reflects the dominance of harmonic interactions in heat transport across 
short-ordered systems, an observation that is not surprising in light of previous 
computations \cite{Segal03,Leitner16} and experiments \cite{Dlott,Gotsmann,Rubtsovalkane,RubtsovAcc}. 

Theoretical and experimental works agree that heat propagates elastically-ballistically
 in short alkanes with 5-20 units. 
Therefore, harmonic-Landauer simulations are typically appropriate for alkane chains.
However, 
other families of molecules such as polyethylene glycol oligomers show fast internal
thermalization and some signatures of diffusive motion \cite{RubtsovPEG,Leitner17}. 
The QSCR method is suitable to capture the elastic
to inelastic (coherent to diffusive) transition in such systems. 
The built-in evaluation of local site temperature in the QSCR method
provides a direct measure to the notion of local temperature and the development of
thermalization in the junction.
In large disordered molecules such as proteins, DNA, and amorphous polymers,
phonon scattering due to anharmonic interactions is expected to be influential to thermal transport 
\cite{LeitnerRev1}, and it is therefore interesting to explore the applicability of the QSCR method to such complex 
systems.

In closing, complementing rigorous approaches, the self-consistent reservoir method
implements the consequences of genuine vibrational anharmonicity 
in a phenomenological manner.
Likewise, the electronic QSCR method, termed B\"uttiker's probe technique \cite{Buttiker}, 
emulates genuine many body (electron-electron, electron-phonon) scattering effects
via electronic probes.
B\"uttiker's probe method has been recently used
to perform massive charge transport simulations in DNA junctions  
\cite{Kilgour1,Roman2}. 
Similarly, future work will be focused on the implementation of the QSCR method in 
combination with atomistic Green's function techniques. This would allow one
to examine the survival of quantum coherent effects in single-molecule phononic conductance 
\cite{Pauli,Gemma1,Cunib1,Cuevas17,Cuevas}, as well as to
systematically explore nanostructured materials with improved properties, e.g. 
superior high, or exceptionally poor thermal transport characteristics.

%
%
\begin{acknowledgments}
D.S. acknowledges the Natural Sciences and Engineering Research Council
of Canada Discovery Grant and the Canada Research Chairs Program.
\end{acknowledgments}



\vspace{5mm}
\renewcommand{\theequation}{A\arabic{equation}}
\setcounter{equation}{0}  
\section*{APPENDIX A: The Landauer Formula}
Equation (\ref{eq:current}) can be organized in the standard language of the phononic Green's 
function and reservoirs' self energies by defining the spectral function of each bath as
 $\tilde \Gamma_s(\omega)$, which is matrix with the dimension of $N\times N$.
This $s$-labeled matrix
includes a single nonzero element at the $s$ row on the diagonal, 
$\left[\tilde\Gamma_s(\omega)\right]_{s,s}=2\omega\Gamma_s(\omega)$.
With this notation, the retarded and advanced Greens' function are identified by
$ G^r(\omega)=\left[\omega^2 I - K_s+i\sum_s \tilde \Gamma_s(\omega)/2\right]^{-1}$,  
with $I$ as the identity matrix and $G_a(\omega)=(G_r(\omega))^{\dagger}$.
Going back to Eq. (\ref{eq:current}) we organize it as
%
\bea
F_s=
\frac{1}{4\pi}\sum_{m=1}^{N}\int_{-\infty}^{\infty}d\omega\, \mathcal T_{s,m}(\omega)\hbar \omega
[f(\omega, T_s)-f(\omega, T_m)],
\nonumber\\
\label{eq:currentA}
\eea
or
\bea
F_s=
\frac{1}{2\pi}\sum_{m=1}^{N}\int_{0}^{\infty}d\omega\, \mathcal T_{s,m}(\omega)\hbar \omega
[f(\omega, T_s)-f(\omega, T_m)].
\nonumber\\
\label{eq:currentB}
\eea
Here the transmission function between two terminals is given by
 $\mathcal T_{s,m}(\omega) =  {\rm Tr}[\tilde \Gamma_s(\omega) G_r(\omega)\tilde\Gamma_m(\omega)G_a(\omega)]$.
When $\gamma_I=0$, which translates to
$[\Gamma_s(\omega)]_{s,s}=0$ for $s=2,3,..,N-1$, 
Equation (\ref{eq:currentB})  immediately reduces to the well known Landauer-like formula for phononic conduction 
\cite{Kirczenow,Segal03,DharRev}, 
showing only the direct left-to-right ($1$ to $N$) transmission function,
\bea
F_1^H=
\frac{1}{2\pi}\int_{0}^{\infty}d\omega\, \mathcal T_{1,N}(\omega)\hbar \omega
[f(\omega, T_1)-f(\omega, T_N)].
\nonumber\\
\label{eq:currentC}
\eea
Note that the self energy can be defined beyond the ohmic limit, as implemented e.g. 
in Ref. \cite{nanoSCR} for the QSCR method.
Equation (\ref{eq:currentC}) can be derived systematically from the nonequilibrium Green's function technique 
\cite{BijayRev,SegalRev}. It has been used extensively in
phononic heat conduction simulations in combination with frequencies and coupling constants obtained 
from density functional theory calculations, see e.g. Refs. \cite{Pauli,Gemma1,Cuevas17,Cuevas}.


\vspace{5mm}
\renewcommand{\theequation}{B\arabic{equation}}
\setcounter{equation}{0}  
\section*{APPENDIX B:  Temperature dependence of the heat current}

\begin{figure*}
\centering
\begin{subfigure}{.5\textwidth}
\centering
\includegraphics[width=8.0cm]{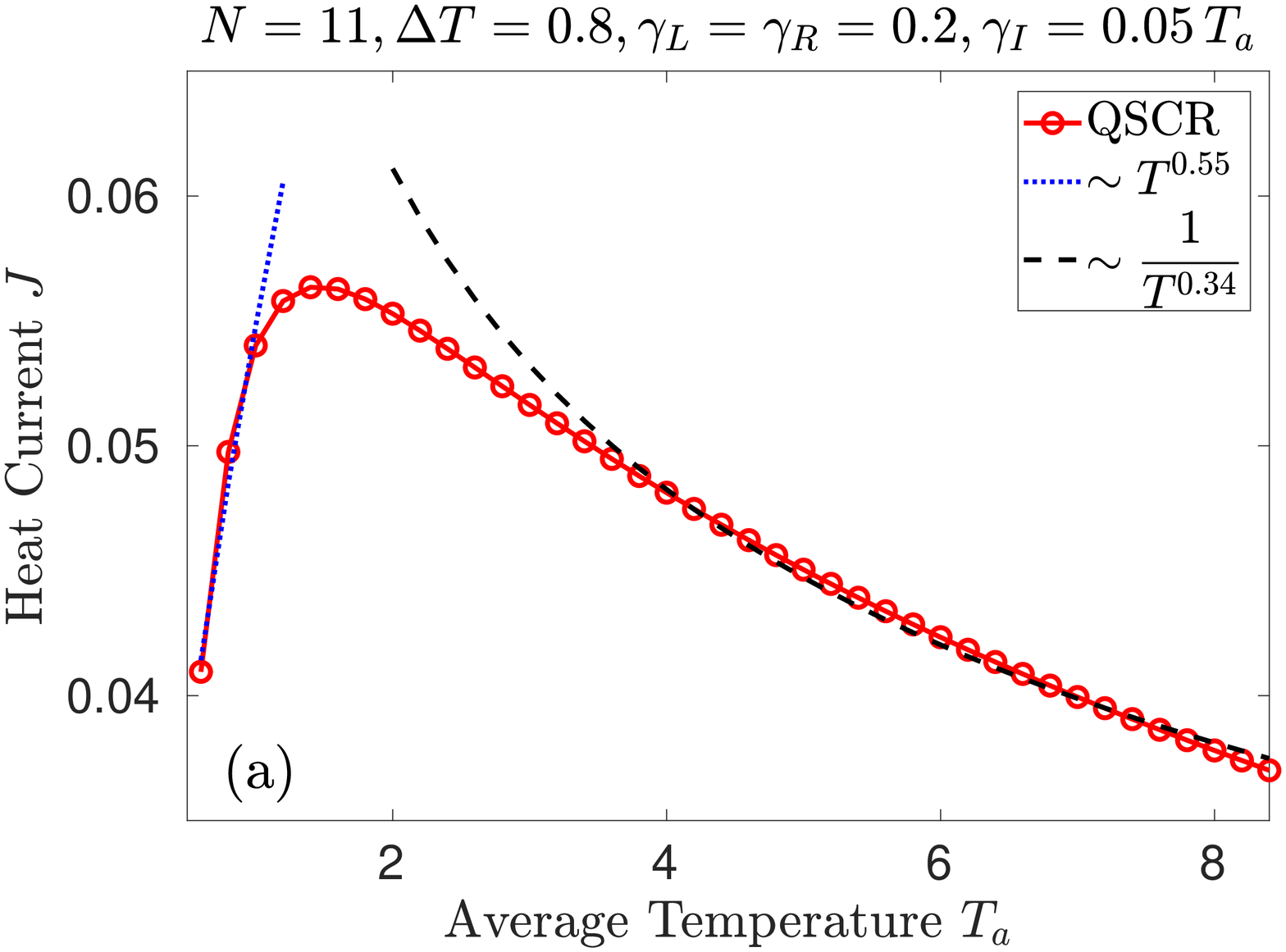} 
\end{subfigure}%
\begin{subfigure}{.5\textwidth}
\centering 
\includegraphics[width=8.0cm]{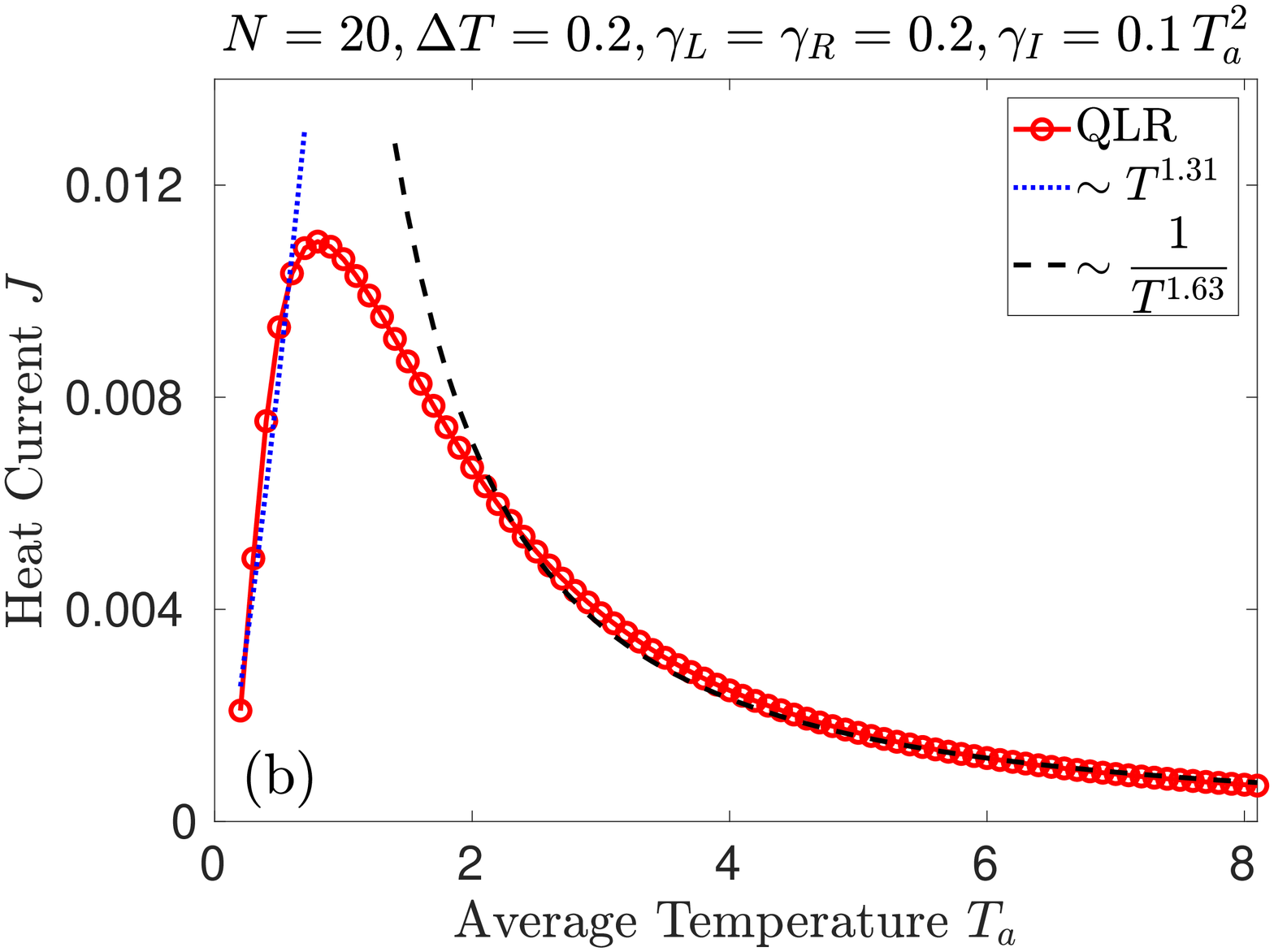}  
\end{subfigure}
\caption{
Heat current vs. average temperature using the (a) QSCR method (circle) with $\gamma_I(T_a)\propto T_a$,
 (b) QLR calculation (circle) with $\gamma_I(T_a)\propto T_a^2$.
Dotted and dashed lines represent the best fit at low and high temperatures, respectively.}
\label{Fig6}
\end{figure*}

Complementing Fig. \ref{Fig5}, we demonstrate here that 
 different scaling for the heat current with $T_a$ can be reached, by
modifying the temperature dependence of the scattering rate $\gamma_I(T_a)$.
We examine a linear dependence,  $\gamma_I(T_a)\propto T_a$, in Fig. \ref{Fig6}(a), and
a quadratic form, $\gamma_I(T_a)\propto T_a^2$, in Fig. \ref{Fig6}(b).
This latter choice in fact more physically
emulates phonon-phonon collision processes, as the occupation factor of each incoming phonon  scales with temperature \cite{Straub}. 
While both models demonstrate a turnover behavior of the current with temperature around the same point,
they show distinct scaling laws. 
Therefore, by adjusting the function $\gamma_I(T_a)$, which physically corresponds to the IVR rate, 
we can capture different types of scattering processes.
Note that in Fig. \ref{Fig6}(b) we employ the QLR method given the relatively small 
temperature difference assumed.

\end{document}